\documentclass[onecolumn,showpacs,preprintnumbers]{revtex4}
\usepackage{graphicx}
\usepackage{dcolumn}
\usepackage{bm}
\usepackage{epsfig}
\usepackage{amsmath,amssymb}

\begin{document}

\title{Analysis of the heavy tensor meson's strong decay with QCD sum rules }
\author{Zhen-Yu Li$^{1}$}
\email{zhenyvli@163.com }
\author{Zhi-Gang Wang$^{2}$ }
\email{zgwang@aliyun.com }
\author{Guo-Liang Yu$^{2}$}
\affiliation{$^1$ School of Physics and Electronic Science, Guizhou Normal College, Guiyang 550018,
China\\$^2$ Department of Mathematics and Physics, North China Electric power university, Baoding 071003,
China}
\date{\today }

\begin{abstract}
In this article, the tensor-vector-pseudoscalar type of vertex is analyzed with the QCD sum rules and the local-QCD sum rules. The hadronic coupling constants $G_{D_{2}^{*}D^{*}\pi}$, $G_{D_{s2}^{*}D^{*}K}$, $G_{B_{2}^{*}B^{*}\pi}$ and $G_{B_{s2}^{*}B^{*}K}$, and the corresponding decay widths are calculated. The results indicate that the QCD sum rules and the local-QCD sum rules give the consistent descriptions. Finally, the full widths of $\Gamma(D_{2}^{*}(2460)^{0})$, $\Gamma(D_{s2}^{*}(2573))$, $\Gamma(B_{2}^{*}(5747)^{0})$ and $\Gamma(B_{s2}^{*}(5840)^{0})$ are discussed in detail.

\end{abstract}


\maketitle


\section*{I. INTRODUCTION}

\label{sec1}

In recent years, many charmonium-like states and several heavy meson's excited states  have been observed (or confirmed) by some collaborations~\cite{ESS,Ber,KAO}. Especially, the masses and the widths of some high exited mesons are obtained experimentally~\cite{VMA,KAO}, such as $D_{2}^{*}(2460)^{0}$,  $D_{s2}^{*}(2573)$,  $B_{2}^{*}(5747)^{0}$, and  $B_{s2}^{*}(5840)^{0}$. The quantum numbers $I(J^{P})$ for $D_{2}^{*}(2460)^{0}$, $B_{2}^{*}(5747)^{0}$ and $B_{s2}^{*}(5840)^{0}$  are $\frac{1}{2}(2^{+})$, $\frac{1}{2}(2^{+})$ and $0(2^{+})$, respectively. As for $D_{s2}^{*}(2573)$, the quantum numbers $I(J^{P})$ = $0(2^{+})$  are favored by the width and decay modes from experiments. Now, these four states have been  included into the $J_{s_{l}}^{P}$ = $(1^{+}, 2^{+})_{\frac{3}{2}}$ doublets, where $s_{l}$ denotes the total angular momentum of the light antiquark in heavy-light mesons~\cite{Ber,KAO}.

 The QCD sum rules (QCDSR) is a powerful non-perturbative theoretical tool in studying the ground state hadrons, and has been widely used in describing the masses, decay constants, hadronic form-factors, and hadronic coupling constants, etc~\cite{MAS,LJR,SNa}.
  Recently, Wang et al.~\cite{Wang1} studied the masses and the decay constants of $D_{2}^{*}(2460)^{0}$,  $D_{s2}^{*}(2573)$,  $B_{2}^{*}(5747)^{0}$, and $B_{s2}^{*}(5840)^{0}$ with the QCDSR, postulating these states are the tensor mesons. Consequently, considering the initial state as a tensor meson and the final states as two pseudoscalar mesons with the three-point QCDSR, K. Azizi et al.~\cite{Azizi} analyzed the strong dacay $D_{2}^{*}(2460)^{0}\rightarrow D^{+}\pi^{-}$ and $D_{s2}^{*}(2573)^{+}\rightarrow D^{+}K^{0}$. And Wang~\cite{Wang2} analyzed the hadronic coupling constants and the decay widths for $D_{2}^{*}(2460)^{0}\rightarrow D\pi$, $D_{s2}^{*}(2573)\rightarrow DK$, $B_{2}^{*}(5747)^{0}\rightarrow B\pi$, and $B_{s2}^{*}(5840)^{0}\rightarrow BK$, and discussed the full widths of $D_{2}^{*}(2460)^{0}$,  $D_{s2}^{*}(2573)$,  $B_{2}^{*}(5747)^{0}$, and  $B_{s2}^{*}(5840)^{0}$ in detail.

 In this work, we focus on the analysis of the strong decays $D_{2}^{*}(2460)^{0}\rightarrow D^{*}\pi$, $D_{s2}^{*}(2573)\rightarrow D^{*}K$, $B_{2}^{*}(5747)^{0}\rightarrow B^{*}\pi$, and $B_{s2}^{*}(5840)^{0}\rightarrow B^{*}K$ with the QCDSR and the local-QCDSR, so as to improve the works in Ref.~\cite{Wang2} and provide necessary information about the properties of the tensor mesons. It's also meaningful to the relevant experiments from the BESIII, LHCB, CDF, D0, and KEK-B collaborations in the futures.

 This paper is organized as follows. We derive the QCDSR and the local-QCDSR for the hadronic coupling constants in the vertices $D_{2}^{*}D^{*}\pi$,  $D_{s2}^{*}D^{*}K$,  $B_{2}^{*}B^{*}\pi$ and $B_{s2}^{*}B^{*}K$ in
Sect.II; In Sect.III, we present the numerical results and calculate the corresponding two body strong decays. And Sect.IV is reserved for our conclusions.

\section*{II. QCDSR FOR THE HADRONIC COUPLING CONSTANTS}

\label{sec2}

In this work, we analyze the tensor-vector-pseudoscalar ($TVP$) type of vertex, for which the three-points correlation function is written down as follow:
\begin{eqnarray}
&& \Pi_{\mu\nu\tau}(p,p')=i^{2}\int d^{4}xd^{4}ye^{ip'\cdot x}e^{i(p-p')\cdot (y-z)} \langle 0|T\{J_{\tau}(x) J_{\mathbb{P}}(y) J^{\dagger}_{\mu\nu}(z) \}|0\rangle |_{z=0},
\end{eqnarray}
\begin{eqnarray}
&& J_{\tau}(x)= \bar{Q}(x)\gamma_{\tau}q(x),  \notag \\
&&  J_{\mathbb{P}}=\bar{q}(y)i\gamma_{5}q'(y),  \notag \\
&& J_{\mu\nu}(z)=i\bar{Q}(z)(\gamma_{\mu}\overleftrightarrow{D}_{\nu}+
\gamma_{\nu}\overleftrightarrow{D}_{\mu}-\frac{2}{3}\tilde{g}_{\mu\nu}\overleftrightarrow{\not{D}})q'(z),  \notag \\
&& \overleftrightarrow{D}_{\mu}=(\overrightarrow{\partial}_{\mu}-ig_{s}G_{\mu})-(\overleftarrow{\partial}_{\mu}+ig_{s}G_{\mu}), \notag \\
&& \tilde{g}_{\mu\nu}=g_{\mu\nu}-\frac{p_{\mu}p_{\nu}}{p^{2}},
\end{eqnarray}
where $Q=c,b$ and $q,q'=u,d,s$, the vector currents $J_{\tau}(x)$ interpolate the heavy vector mesons $D^{*}$ or $B^{*}$,  the pseudoscalar currents $j_{\mathbb{P}}(y)$ interpolate the light pseudoscalar mesons $\pi$ or $K$, and the tensor currents $J_{\mu\nu}(z)$ interpolate the heavy tensor mesons $D_{2}^{*}(2460)^{0}$,  $D_{s2}^{*}(2573)$,  $B_{2}^{*}(5747)^{0}$, and  $B_{s2}^{*}(5840)^{0}$, respectively.
\subsection*{1. THE HADRONIC SIDE}

\label{sec1}
With the same quantum numbers as the current operators $J_{\mu\nu}(z)$, $J_{\tau}(x)$ and $j_{\mathbb{P}}(y)$, a complete set of intermediate hadronic states are inserted into the correlation functions $\Pi_{\mu\nu\tau}(p,p')$. So that, the hadronic representation is obtained~\cite{MAS,LJR}. After isolating the ground state contributions from the heavy tensor mesons $\mathbb{T}$, heavy vector mesons $\mathbb{V}$ and light pseudoscalar mesons $\mathbb{P}$, the correlation function is expressed as
\begin{equation}\label{sdgfergn}
\begin{aligned}
\Pi_{\mu\nu\tau}^{HAD}(p,p')
&=\frac{f_{\mathbb{P}}M_{\mathbb{P}}^{2}f_{\mathbb{V}}M_{\mathbb{V}}f_{\mathbb{T}}
M_{\mathbb{T}}^{2}G_{\mathbb{TVP}}}{(m_{Q}+m_{q})(M_{\mathbb{V}}^{2}-p'^{2})(M_{\mathbb{P}}^{2}-q^{2})(M_{\mathbb{T}}^{2}-p^{2})}\\
&\{\frac{1}{2}(p'^{\mu}\epsilon^{\nu\tau pp'}+p'^{\nu}\epsilon^{\mu\tau pp'})+(\frac{-p^{2}-p'^{2}+q^{2}}{4p^{2}})
(p^{\mu}\epsilon^{\nu\tau pp'}+p^{\nu}\epsilon^{\mu\tau pp'})\}+\cdot\cdot\cdot\\
&=\Pi_{1}^{had}(p^{2},p'^{2})p'^{\mu}\epsilon^{\nu\tau pp'}+\Pi_{2}^{had}(p^{2},p'^{2})p'^{\nu}\epsilon^{\mu\tau pp'}\\
&+\Pi_{3}^{had}(p^{2},p'^{2})p^{\mu}\epsilon^{\nu\tau pp'}+\Pi_{4}^{had}(p^{2},p'^{2})p^{\nu}\epsilon^{\mu\tau pp'}+\cdot\cdot\cdot,
\end{aligned}
\end{equation}
where the decay constants $f_{\mathbb{T}}$, $f_{\mathbb{V}}$, $f_{\mathbb{P}}$ and the hadronic coupling constants $G_{\mathbb{TVP}}$ are defined by
\begin{eqnarray}
&& \langle0|J_{\mu\nu}(0)|\mathbb{T}(p)\rangle\equiv f_{\mathbb{T}}M_{\mathbb{T}}^{2}\xi_{\mu\nu}^{*}(s,p), \notag \\
&& \langle0|J_{\tau}(0)|\mathbb{V}(p')\rangle\equiv f_{\mathbb{V}}M_{\mathbb{V}}\zeta_{\tau}^{*}(p'), \notag \\
&& \langle0|J_{\mathbb{P}}(0)|\mathbb{P}(q)\rangle\equiv f_{\mathbb{P}}M_{\mathbb{P}}^{2}/(m_{Q}+m_{q}),
\end{eqnarray}
\begin{equation}
\langle\mathbb{V}(p')\mathbb{P}(q)|\mathbb{T}(p)\rangle=G_{\mathbb{TVP}}\epsilon^{\alpha\beta\omega\rho}
p_{\alpha}\xi_{\beta\eta}^{(\lambda)}q^{\eta}p'_{\omega}\zeta_{\rho},  \label{Usop}
\end{equation}
with $q=p-p'$, and the $\xi_{\mu\nu}$ are the polarization vectors of the tensor mesons with the following properties,
\begin{eqnarray}
\sum_{s}\xi_{\mu\nu}^{*}(s,p)\xi_{\alpha\beta}(s,p)
=\frac{\tilde{g}_{\mu\alpha}\tilde{g}_{\nu\beta}+\tilde{g}_{\mu\beta}
\tilde{g}_{\nu\alpha}}{2}-\frac{\tilde{g}_{\mu\nu}\tilde{g}_{\alpha\beta}}{3}. \label{current}
\end{eqnarray}

\subsection*{2. THE OPE SIDE}

\label{sec2}
After contracting the quark fields with Wick theorem in Eq.(1), the correlation function is written down as follow:
\begin{eqnarray}
\Pi_{\mu\nu\tau}^{OPE}(p,p')=-i^{4}\int d^{4}xd^{4}ye^{ip'\cdot x}e^{i(p-p')\cdot (y-z)}
Tr\{\gamma_{\tau}S_{mn}^{q}(x-y)\gamma_{5}S_{nk}^{q'}(y-z)\Gamma_{\mu\nu}S_{km}^{Q}(z-x)\}  |_{z=0},
\end{eqnarray}
where
\begin{eqnarray}
\Gamma_{\mu\nu}=i(\gamma_{\mu}\frac{\overleftrightarrow{\partial}}{\partial z^{\nu}}+\gamma_{\nu}
\frac{\overleftrightarrow{\partial}}{\partial z^{\mu}}-\frac{2}{3}\tilde{g}_{\mu\nu}\gamma^{\omega}
\frac{\overleftrightarrow{\partial}}{\partial z^{\omega}}),
\end{eqnarray}

\begin{equation}\label{shhgn}
\begin{aligned}
S_{ij}^{Q}(x)&=\frac{i}{(2\pi)^{4}}\int d^{4}ke^{-ik\cdot x}\{\frac{\delta_{ij}}{\not{k}-m_{Q}}
-\frac{g_{s}G_{\alpha\beta}^{n}t_{ij}^{n}}{4}\frac{\sigma^{\alpha\beta}(\not{k}+m_{Q})+(\not{k}+m_{Q})\sigma^{\alpha\beta}}{(k^{2}-m_{Q}^{2})^{2}}\\
&+\frac{g_{s}^{2}(t^{a}t^{b})_{ij}G_{\alpha\beta}^{a}G_{\mu\nu}^{b}(f^{\alpha\beta\mu\nu}+f^{\alpha\mu\beta\nu}+f^{\alpha\mu\nu\beta})}
{4(k^{2}-m_{Q}^{2})^{5}}\\
& +\frac{i\langle g_{s}^{3}GGG\rangle}{48}\frac{(\not{k}+m_{Q})[\not{k}(k^{2}-3m_{Q}^{2})+2m_{Q}(2k^{2}-m_{Q}^{2})](\not{k}+m_{Q})}{(k^{2}-m_{Q}^{2})^{6}}+\cdot\cdot\cdot\},\\
\end{aligned}
\end{equation}
\begin{eqnarray}
&& f^{\alpha\beta\mu\nu}=(\not{k}+m_{Q})\gamma^{\alpha}(\not{k}+m_{Q})\gamma^{\beta}(\not{k}+m_{Q})\gamma^{\mu}(\not{k}+m_{Q})\gamma^{\nu}(\not{k}+m_{Q}),\notag
\end{eqnarray}
$t^{n}=\frac{\lambda^{n}}{2}$, the $\lambda^{n}$ is the Gell-Man matrix, the $i,j,k$ are color indices~\cite{LJR}. In common, the full light quark propagators are chosen in the coordinate space. In the present case, the quark condensates and mixed condensates have no contributions (see Ref.~\cite{GLY15}), so the full $q/q'$ quark propagators are obtained with a simple replacement $Q\rightarrow q/q'$. In addition, the gluon field $G_{\mu}(z)$ in the covariant derivative has no contributions as $G_{\mu}(z)=\frac{1}{2}z^{\lambda}G_{\lambda\mu}(0)+\cdot\cdot\cdot=0$.

Firstly, according to Eqs.(7), (8) and (9), the leading-order contributions can be written as
\begin{equation}\label{shhgn}
\begin{aligned}
\Pi_{\mu\nu\tau}^{0}(p,p')&=\frac{3}{(2\pi)^{4}}\int d^{4}k\frac{Tr\{\gamma_{\tau}[\not{k}+m_{q}]\gamma_{5}[\not{k}+\not{p}-\not{p'}+m_{q'}]\Gamma_{\mu\nu}[\not{k}-\not{p'}+m_{Q}]\}}
{[k^{2}-m_{q}^{2}][(k+p-p')^{2}-m_{q'}^{2}][(k-p')^{2}-m_{Q}^{2}]}\\
&=\int dsdu\frac{\rho_{\mu\nu\tau}^{0}(s,u,q^{2})}{(s-p^{2})(u-p'^{2})},
\end{aligned}
\end{equation}
where
\begin{eqnarray}
\Gamma_{\mu\nu}=\gamma_{\mu}(2k+p-2p')_{\nu}+\gamma_{\nu}(2k+p-2p')_{\mu}-\frac{2}{3}\tilde{g}_{\mu\nu}(2\not{k}+\not{p}-2\not{p'}).
\end{eqnarray}
Using the Cutkosky's rules (see Fig.1), the quark lines are put on mass-shell, and the leading-order spectral densities $\rho_{\mu\nu\tau}^{0}$ is obtained as follow,
\begin{equation}\label{sllgn}
\begin{aligned}
\rho_{\mu\nu\tau}^{0}(s,u,q^{2})&=\frac{3}{4\pi^{2}\sqrt{\lambda}}\{(2B-2)[(m_{q}-m_{Q})B+(m_{Q}-m_{q'})A+m_{q}](p'^{\mu}\epsilon^{\nu\tau pp'}+p'^{\nu}\epsilon^{\mu\tau pp'})\\
&+(2A+1)[(m_{q}-m_{Q})B+(m_{Q}-m_{q'})A+m_{q}](p^{\mu}\epsilon^{\nu\tau pp'}+p^{\nu}\epsilon^{\mu\tau pp'})\}   \\
&\equiv\rho_{1}^{0}p'^{\mu}\epsilon^{\nu\tau pp'}+\rho_{2}^{0}p'^{\nu}\epsilon^{\mu\tau pp'}
+\rho_{3}^{0}p^{\mu}\epsilon^{\nu\tau pp'}+\rho_{4}^{0}p^{\nu}\epsilon^{\mu\tau pp'},
\end{aligned}
\end{equation}
where
\begin{eqnarray}
&\lambda\equiv\lambda(s,u,q^{2})=(s+u-q^{2})^{2}-4su , \notag \\
&A\equiv A(s,u,q^{2})=\frac{(u+m_{q}^{2}-m_{Q}^{2})(s+u-q^{2})-2u(u-q^{2}+m_{q'}^{2}-m_{Q}^{2})}{\lambda(s,u,q^{2})} , \label{qq} \\
&B\equiv B(s,u,q^{2})=\frac{(u-q^{2}+m_{q'}^{2}-m_{Q}^{2})(s+u-q^{2})-2s(u+m_{q}^{2}-m_{Q}^{2})}{\lambda(s,u,q^{2})} . \notag
\end{eqnarray}
\begin{figure}[htbp]
\begin{center}
\includegraphics[width=0.4\textwidth]{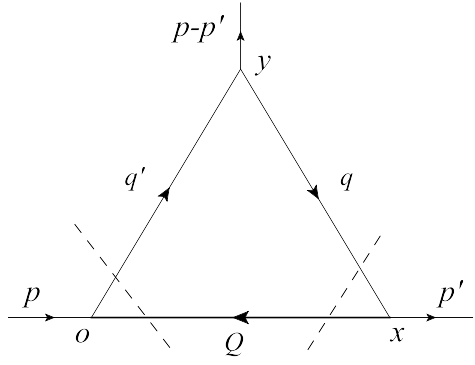}
\end{center}
\caption{The leading order contributions, the dashed lines denotes the Cutkosky's cuts. }
\end{figure}

In the next step, the vacuum condensates contributions up to dimension-6, i.e. the gluon condensates of $\langle GG\rangle$ and $\langle GGG\rangle$ in present work, are considered (see Figs.2 and 3). The corresponding results are arranged in the appendix A and B.

At last, the correlation function in the OPE side is combined as follow:
\begin{equation}\label{sfegn}
\begin{aligned}
\Pi_{\mu\nu\tau}^{OPE}(p,p')&=\Pi_{\mu\nu\tau}^{0}(p,p')+\Pi_{\mu\nu\tau}^{\langle GG\rangle}(p,p')+\Pi_{\mu\nu\tau}^{\langle GGG\rangle}(p,p')\\
&=[\Pi_{1}^{0}(p^{2},p'^{2},q^{2})+\Pi_{1}^{\langle GG\rangle}(p^{2},p'^{2},q^{2})+\Pi_{1}^{\langle GGG\rangle}(p^{2},p'^{2},q^{2})]p'^{\mu}\epsilon^{\nu\tau pp'}\\
&+[\Pi_{2}^{0}(p^{2},p'^{2},q^{2})+\Pi_{2}^{\langle GG\rangle}(p^{2},p'^{2},q^{2})+\Pi_{2}^{\langle GGG\rangle}(p^{2},p'^{2},q^{2})]p'^{\nu}\epsilon^{\mu\tau pp'}\\
&+[\Pi_{3}^{0}(p^{2},p'^{2},q^{2})+\Pi_{3}^{\langle GG\rangle}(p^{2},p'^{2},q^{2})+\Pi_{3}^{\langle GGG\rangle}(p^{2},p'^{2},q^{2})]p^{\mu}\epsilon^{\nu\tau pp'}\\
&+[\Pi_{4}^{0}(p^{2},p'^{2},q^{2})+\Pi_{4}^{\langle GG\rangle}(p^{2},p'^{2},q^{2})+\Pi_{4}^{\langle GGG\rangle}(p^{2},p'^{2},q^{2})]p^{\nu}\epsilon^{\mu\tau pp'}.
\end{aligned}
\end{equation}
\begin{figure}[htbp]
\begin{center}
\includegraphics[width=0.8\textwidth]{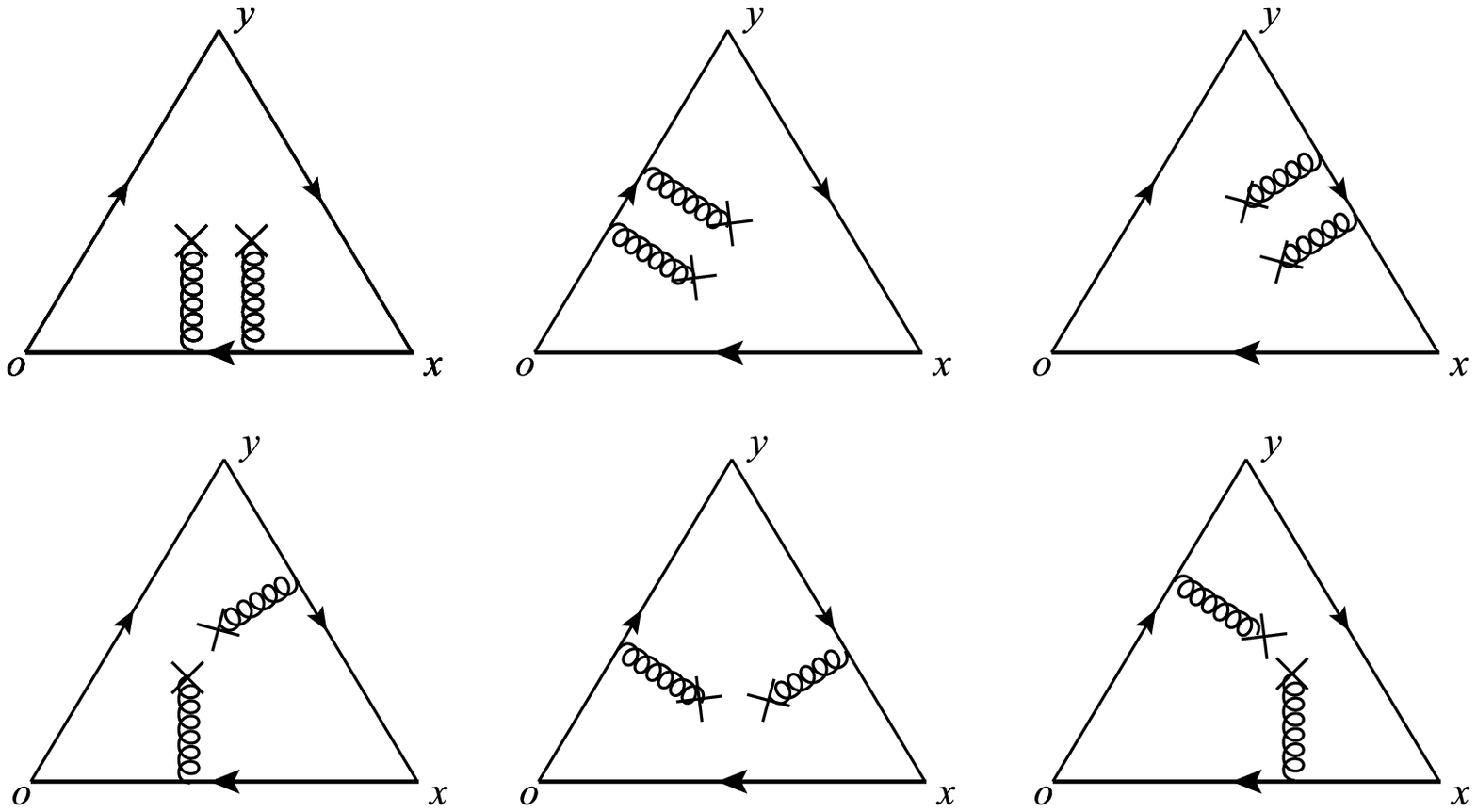}
\end{center}
\caption{The $\langle GG\rangle$ condensates contributions}
\end{figure}
\begin{figure}[htbp]
\begin{center}
\includegraphics[width=0.8\textwidth]{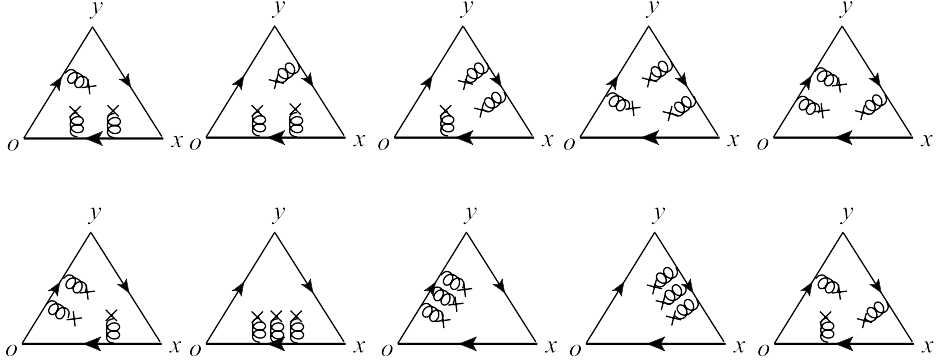}
\end{center}
\caption{The $\langle GGG\rangle$ condensates contributions }
\end{figure}

\subsection*{3. THE QCDSR}

\label{sec3}
Changing the variables into the Euclidean space, i.e. $p^{2}\rightarrow -P^{2}$, $p'^{2}\rightarrow -P'^{2}$ and $q^{2}\rightarrow -Q^{2}$, performing the double Borel transformation, and taking quark-hadron duality below the continuum thresholds $s_{0}$ and $u_{0}$ respectively, the QCDSR are obtained as follows:
\begin{equation}\label{sddgn}
\begin{aligned}
\hat{\Pi}_{i}(M_{1}^{2},M_{2}^{2},Q^{2})&=\frac{1}{2}\frac{f_{\mathbb{P}}M_{\mathbb{P}}^{2}f_{\mathbb{V}}M_{\mathbb{V}}f_{\mathbb{T}}
M_{\mathbb{T}}^{2}G_{\mathbb{TVP}}}{(m_{Q}+m_{q})(M_{\mathbb{P}}^{2}+Q^{2})}\frac{1}{M_{1}^{2}M_{2}^{2}}
e^{-\frac{M_{\mathbb{T}}^{2}}{M_{1}^{2}}}e^{-\frac{M_{\mathbb{V}}^{2}}{M_{2}^{2}}}\\
&=\frac{1}{M_{1}^{2}M_{2}^{2}}\int dsdue^{-\frac{s}{M_{1}^{2}}}e^{-\frac{u}{M_{2}^{2}}}\hat{\rho}_{i}^{0}(s,u,Q^2)\\
&+\hat{\Pi}_{i}^{\langle GG\rangle}(M_{1}^{2},M_{2}^{2},Q^{2})+\hat{\Pi}_{i}^{\langle GGG\rangle}(M_{1}^{2},M_{2}^{2},Q^{2}),
\end{aligned}
\end{equation}
\begin{equation}\label{sldgn}
\begin{aligned}
\hat{\Pi}_{j}(M_{1}^{2},M_{2}^{2},Q^{2})&=\frac{1}{4}\frac{f_{\mathbb{P}}M_{\mathbb{P}}^{2}f_{\mathbb{V}}M_{\mathbb{V}}f_{\mathbb{T}}
M_{\mathbb{T}}^{2}G_{\mathbb{TVP}}}{(m_{Q}+m_{q})(M_{\mathbb{P}}^{2}+Q^{2})}\frac{1}{M_{1}^{2}M_{2}^{2}}
e^{-\frac{M_{\mathbb{V}}^{2}}{M_{2}^{2}}}(\frac{M_{\mathbb{V}}^{2}+Q^{2}}{M_{\mathbb{T}}^{2}}
+\frac{M_{\mathbb{V}}^{2}+Q^{2}-M_{\mathbb{T}}^{2}}{M_{\mathbb{T}}^{2}}e^{-\frac{M_{\mathbb{T}}^{2}}{M_{1}^{2}}})\\
&=\frac{1}{M_{1}^{2}M_{2}^{2}}\int dsdue^{-\frac{s}{M_{1}^{2}}}e^{-\frac{u}{M_{2}^{2}}}\hat{\rho}_{j}^{0}(s,u,Q^2)\\
&+\hat{\Pi}_{j}^{\langle GG\rangle}(M_{1}^{2},M_{2}^{2},Q^{2})+\hat{\Pi}_{j}^{\langle GGG\rangle}(M_{1}^{2},M_{2}^{2},Q^{2}),
\end{aligned}
\end{equation}
where $M_{1}^{2}$ and $M_{2}^{2}$ are Borel parameters. $i=1,2$ denote $p'^{\mu}\epsilon^{\nu\tau pp'}$ and $p'^{\nu}\epsilon^{\mu\tau pp'}$ structures,
and $j=3,4$ denote $p^{\mu}\epsilon^{\nu\tau pp'}$ and $p^{\nu}\epsilon^{\mu\tau pp'}$ structures, respectively.

\begin{eqnarray}
&& \int dsdu=\int_{s_{1}}^{s_{0}}ds\int_{u_{1}}^{u_{0}}du|_{|f|\leq1}, \notag \\
&& f=\frac{2s(m_{q}^{2}-m_{Q}^{2}+u)+(m_{Q}^{2}-m_{q'}^{2}-u-Q^2)(s+u+Q^{2})}{\sqrt{(m_{Q}^{2}-m_{q'}^{2}-u-Q^2)^{2}-4sm_{q}^{2}}\sqrt{\lambda(s,u,-Q^{2})}},
\end{eqnarray}
the $s_{1}$ and $u_{1}$ will get their assignments in the next section.

\subsection*{4. THE LOCAL-QCDSR}

\label{sec3}

In this article, the local-QCDSR has also been used, so as to get more comprehensive analysis. Thus, the full spectral density $\rho$, including both the leading order term and the vacuum condensates, are deduced as well.  The result shows that the gluon condensates of $\langle GG\rangle$ and $\langle GGG\rangle$ have no contributions to the full spectral density $\rho$, i.e.  $\rho=\rho^{0}$.

Now the local limit $M_{1}^{2}=M_{2}^{2}\rightarrow\infty$ is taken, and the local-QCDSR are obtained as follows:
\begin{equation}\label{sddgn}
\begin{aligned}
\hat{\Pi}_{i}(Q^{2})&=\frac{1}{2}\frac{f_{\mathbb{P}}M_{\mathbb{P}}^{2}f_{\mathbb{V}}M_{\mathbb{V}}f_{\mathbb{T}}
M_{\mathbb{T}}^{2}G_{\mathbb{TVP}}}{(m_{Q}+m_{q})(M_{\mathbb{P}}^{2}+Q^{2})}
=\int dsdu\hat{\rho}_{i}^{0}(s,u,Q^2),
\end{aligned}
\end{equation}
\begin{equation}\label{skkgn}
\begin{aligned}
\hat{\Pi}_{j}(Q^{2})&=\frac{1}{4}\frac{f_{\mathbb{P}}M_{\mathbb{P}}^{2}f_{\mathbb{V}}M_{\mathbb{V}}f_{\mathbb{T}}
M_{\mathbb{T}}^{2}G_{\mathbb{TVP}}}{(m_{Q}+m_{q})(M_{\mathbb{P}}^{2}+Q^{2})}(2\frac{M_{\mathbb{V}}^{2}+Q^{2}}{M_{\mathbb{T}}^{2}}-1)
=\int dsdu\hat{\rho}_{j}^{0}(s,u,Q^2),
\end{aligned}
\end{equation}
with $i=1,2$ and $j=3,4$.

\section*{III. NUMERICAL RESULTS AND DISCUSSIONS}

In this section, the QCDSR and the local-QCDSR are analyzed numerically for the hadronic coupling constants $G_{\mathbb{TVP}}$. The used input parameters are listed in Table I. The thresholds $s_{0}$, $u_{0}$, $s_{1}$ and $u_{1}$ are evaluated as the following relations~\cite{Wang2},
\begin{equation}\label{soogn}
\begin{aligned}
&s_{0}=(M_{\mathbb{T}}+\triangle M)^2, ~~ u_{0}=(M_{\mathbb{V}}+\triangle M)^2 , \\
&s_{1}=(m_{Q}+m_{q'})^2, ~~ u_{1}=(m_{Q}+m_{q})^2,
\end{aligned}
\end{equation}
where the $\triangle M\equiv0.5\pm0.1GeV$.
\begin{table*}[t]
\begin{ruledtabular}\caption{The used input parameters in this work. }
\begin{tabular}{c c c c c c c c c c c c c c c c c c}
Parameters & \ Values  & \ Parameters  & \ Values    \\
\hline
$m_{u}$ &  \ $2.3_{-0.5}^{+0.7}MeV $~\cite{Ber} & \ $m_{c}$  &  \  $1.275\pm 0.025GeV $~\cite{Ber}   \\
$m_{d}$ &  \ $4.8_{-0.3}^{+0.5}MeV $~\cite{Ber} & \ $m_{b}$  &  \  $4.18\pm 0.03GeV $~\cite{Ber}   \\
$m_{s}$ &  \ $95\pm 5MeV $~\cite{Ber} & \    &  \      \\
\hline
$M_{D_{2}^{*0}}$ &  \ $2462.6\pm0.6MeV $~\cite{Ber} & \ $f_{D_{2}^{*0}}$  &  \  $182 MeV$~\cite{Wang2}   \\
$M_{D_{s2}^{*+}}$ &  \ $2571.9\pm0.8MeV $~\cite{Ber}  & \ $f_{D_{s2}^{*+}}$  & \ $222 MeV$~\cite{Wang2}   \\
$M_{B_{2}^{*0}}$  &  \ $5743\pm5MeV $~\cite{Ber} & \ $f_{B_{2}^{*0}}$  &  \  $110 MeV$~\cite{Wang2}   \\
$M_{B_{s2}^{*0}}$ &  \ $5839.96\pm0.20MeV $~\cite{Ber} & \ $f_{B_{s2}^{*0}}$ &  \  $134 MeV$~\cite{Wang2}    \\
\hline
$M_{D^{*0}}$ &  \ $2006.96\pm0.10MeV $~\cite{Ber} & \ $M_{\pi^{0}}$  &  \  $134.9766\pm 0.0006MeV$~\cite{Ber}    \\
$M_{D^{*+}}$ &  \ $2010.26\pm0.07MeV $~\cite{Ber} & \ $M_{\pi^{+}}$  &  \ $139.57018\pm0.00035MeV $~\cite{Ber}  \\
$f_{D^{*}}$ &  \ $263MeV $~\cite{wang3} & \ $f_{\pi}$  &  \  $130.41MeV $~\cite{Ber}   \\
\hline
$M_{B^{*0}}$ &  \ $5325.2\pm0.4MeV$~\cite{Ber} & \ $M_{K^{0}}$ &  \  $497.614\pm0.024MeV$~\cite{Ber}   \\
$M_{B^{*+}}$ &  \ $5324.25\pm0.44MeV$\cite{Ber}  & \ $M_{K^{+}}$  &  \  $493.677\pm0.016MeV $~\cite{Ber}   \\
$f_{B^{*}}$ &  \ $213MeV$~\cite{wang3} & \ $f_{K}$  &  \ $156.1MeV $~\cite{Ber}  \\
\hline
$\langle \alpha_{s}GG/\pi\rangle$ &  \ $0.022\pm0.004 GeV^{4}$~\cite{SN10} & \ $\langle g_{s}^{3}G^{a}G^{b}G^{c}f^{abc}\rangle$  &  \  $0.616\pm0.385 GeV^{6}$~\cite{SN10}  \\
\end{tabular}
\end{ruledtabular}
\end{table*}

\subsection*{3.1 NUMERICAL RESULTS OF THE QCDSR}

  As to the QCDSR, we observe that the structures $p'^{\mu}\epsilon^{\nu\tau pp'}$ and $p'^{\nu}\epsilon^{\mu\tau pp'}$ are the pertinent structures. Now, we will find the working regions for the auxiliary parameters $M_{1}^{2}$ and $M_{2}^{2}$, of which the hadronic coupling constants $G_{\mathbb{TVP}}$ should roughly be independent. The reasonable working regions for $M_{1}^{2}$ are shown in Fig.4, where the relation
 $M_{1}^{2}/M_{\mathbb{T}}^{2}=M_{2}^{2}/M_{\mathbb{V}}^{2}$ is used~\cite{MEB12}. It is shown in Fig.4, the Borel window $6GeV^{2}\leq M_{1}^{2} \leq 9GeV^{2}$ ($38 GeV^{2}\leq M_{1}^{2} \leq 42 GeV^{2}$) for  $D_{2}^{*}(2460)^{0}$ and $D_{s2}^{*}(2573)$ ($B_{2}^{*}(5747)^{0}$ and $B_{s2}^{*}(5840)^{0}$) is chosen. In these intervals, the dependence of $G_{\mathbb{TVP}}$ on the Borel parameters are weak.

\begin{figure}[htbp]
\begin{center}
\includegraphics[width=1.0\textwidth]{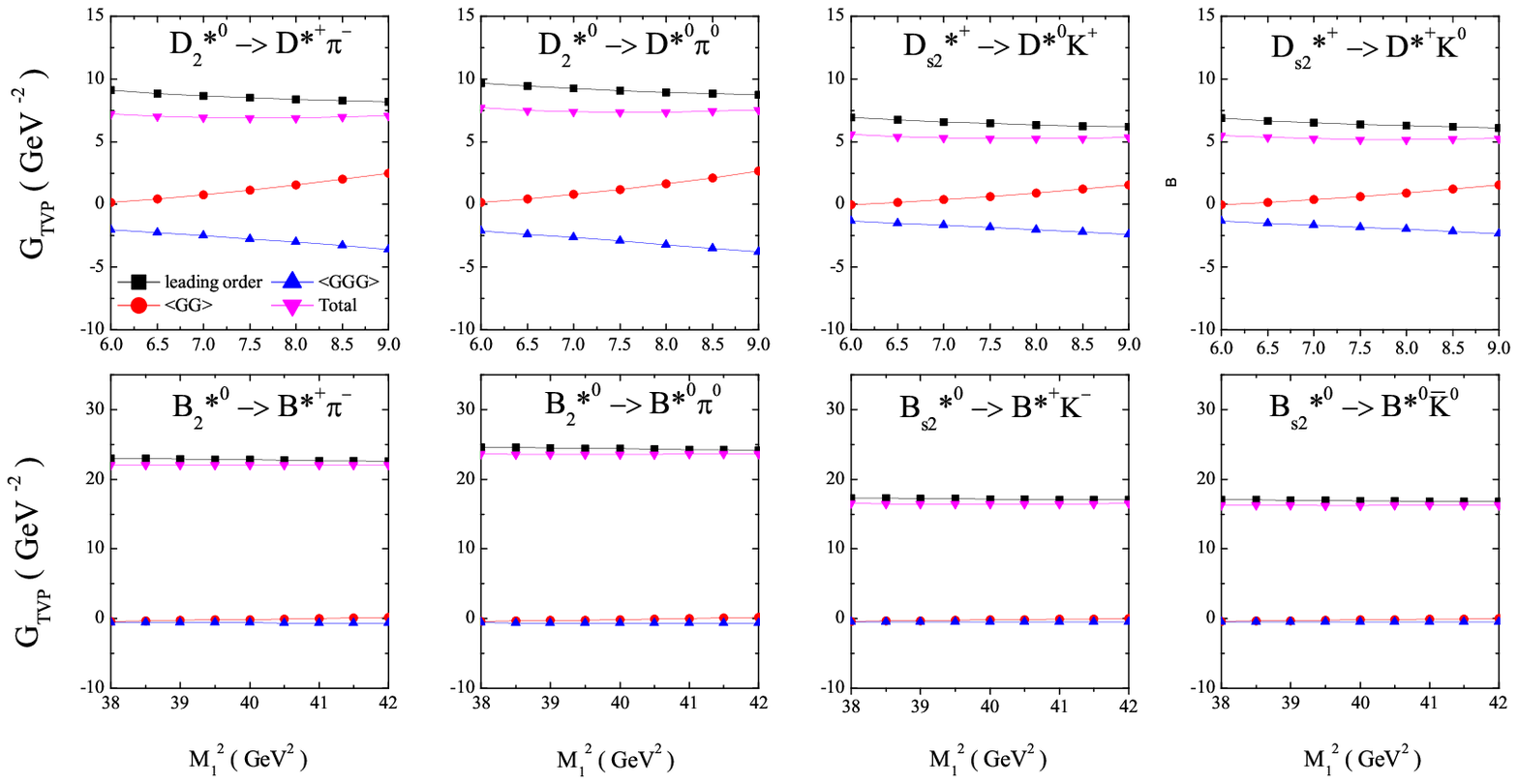}
\end{center}
\caption{(Color online) As to the $p'^{\mu}\epsilon^{\nu\tau p p'}$ or $p'^{\nu}\epsilon^{\mu\tau p p'}$ structure with the QCDSR, $G_{\mathbb{TVP}}$ stability is found for every decay processes with the fixed $Q^{2}=6.5 GeV^{2}$.}
\end{figure}

Then, we proceed to search for the behavior of the hadronic coupling constants $G_{\mathbb{TVP}}$ with respect to $Q^{2}$. Using the method of trials and errors in a large range of the deep Euclidean space, we select out the optimized fitting $Q^{2}$ intervals and the optimal fitting functions $G_{\mathbb{TVP}}(Q^{2})=C_{1}+C_{2}Q^{2}$, and obtain the hadronic coupling constants $G_{\mathbb{TVP}}(Q^{2}=-M_{\mathbb{P}}^{2})$. At the same time, the pole contributions and the leading order contributions are considered as well, and the detailed information are listed in Table II. Generally, the pole contribution should be bigger than the continuum contribution by at least 50\% and that the leading order term contributes with more than 50\% to the total correlation function. From Table II, it is shown that these two indices work well.

In this work, the two-body decay widths can be written as
\begin{equation}\label{sdfgfd}
   \Gamma_{i}=C_{p}\frac{G_{\mathbb{TVP}}^{2}\sqrt{[M_{\mathbb{T}}^{2}-(M_{\mathbb{V}}+M_{\mathbb{P}})^{2}][M_{\mathbb{T}}^{2}-(M_{\mathbb{V}}-M_{\mathbb{P}})^{2}]}
   [(\frac{M_{\mathbb{T}}^{2}-M_{\mathbb{P}}^{2}+M_{\mathbb{V}}^{2}}{2})^{2}-M_{\mathbb{T}}^{2}M_{\mathbb{V}}^{2}]^{2}}{80\pi M_{\mathbb{T}}^{5}},
 \end{equation}
 where $C_{p}$ is equal to 2 (or 1) for $\pi^{\pm}$, $K$ (or $\pi^{0}$), and $i=1,2,3,4$ denotes the individual structure $p'^{\mu}\epsilon^{\nu\tau pp'}$, $p'^{\nu}\epsilon^{\mu\tau pp'}$, $p^{\mu}\epsilon^{\nu\tau pp'}$ and $p^{\nu}\epsilon^{\mu\tau pp'}$, respectively. The numerical results of the decay widths can also be found in Table II. In actual calculations, the uncertainties of the decay constants are neglected so as to avoid doubling counting as the uncertainties originating mainly from the threshold parameters and heavy quark masses. The numerical results of the hadronic coupling constants $G_{\mathbb{TVP}}(Q^{2}=-M_{\mathbb{P}}^{2})$  or the decay widths are of the specific forms with such definition: for example, as to $3.70_{-0.78}^{+1.00}$, $3.70$ is the central value fitted from the central value of the input parameters, and the $3.70+1.00=4.70$ ($3.70-0.78=2.92$) is the upper bound (lower bound) value from the 'upper' ('lower') contribution due to the input parameters' uncertainties.
\begin{table*}[tty]
\begin{ruledtabular}\caption{The numerical results with the QCDSR }
\begin{tabular}{c c c c c c c c c c c c c c c c c c}
$p'^{\mu}\epsilon^{\nu\tau p p'}$($p'^{\nu}\epsilon^{\mu\tau p p'}$) & \  $ Q^{2}  (GeV^{2}) $ & \ Leading Order & \ Pole  &\ $G_{\mathbb{TVP}}(GeV^{-2})$ &\ Decay Width $\Gamma_{i}$ \\
\hline
$ D_{2}^{*0} \rightarrow D^{*+} \pi^{-} $ &\ $3.2\sim4.0$ & \ $72.8\%\sim71.4\%$  & \ $52.6\%\sim46.9\%$ &\ $3.70_{-0.78}^{+1.00}$  &\ $1.97_{-0.75}^{+1.23}MeV$ \\
$ D_{2}^{*0} \rightarrow D^{*0} \pi^{0} $ &\ $3.2\sim4.0$ & \ $72.9\%\sim71.5\%$  & \  $52.5\%\sim46.8\%$ &\ $3.93_{-0.83}^{+1.09}$  &\ $1.18_{-0.45}^{+0.74}MeV$\\
$ D_{s2}^{*+} \rightarrow D^{*0}  K^{+} $ &\ $3.2\sim4.0$ & \ $72.0\%\sim70.8\%$  & \ $54.6\%\sim49.0\%$  &\ $2.77_{-0.47}^{+0.45}$  &\ $102.91_{-41.30}^{+36.12}KeV$\\
$ D_{s2}^{*+} \rightarrow D^{*+}  K^{0} $ &\ $3.2\sim4.0$ & \ $72.0\%\sim70.7\%$   & \ $54.6\%\sim49.1\%$ &\ $2.73_{-0.46}^{+0.45}$  &\ $77.52_{-24.10}^{+27.06}KeV$\\
$ B_{2}^{*0} \rightarrow B^{*+}  \pi^{-} $ &\ $1.7\sim2.5$ &\ $96.0\%\sim96.7\%$  &\ $91.4\%\sim71.2\%$ &\ $2.68_{-0.09}^{+0.04}$  &\ $0.92_{-0.06}^{+0.02}MeV$\\
$ B_{2}^{*0} \rightarrow B^{*0}  \pi^{0} $ & \ $1.7\sim2.5$ & \ $96.0\%\sim96.7\%$  & \ $91.5\%\sim71.3\%$ &\ $2.86_{-0.09}^{+0.05}$  &\ $0.52_{-0.03}^{+0.02}MeV$\\
$ B_{s2}^{*0} \rightarrow B^{*+}  K^{-} $  & \ $1.7\sim2.5$ & \ $94.9\%\sim96.2\%$  & \  $93.0\%\sim72.4\%$ &\ $1.86_{-0.29}^{+0.20}$  &\ $3.23_{-0.94}^{+0.71}KeV$\\
$ B_{s2}^{*0} \rightarrow B^{*0}  \bar{K}^{0} $   & \ $1.7\sim2.5$ & \ $94.9\%\sim96.2\%$  & \  $93.1\%\sim72.4\%$ &\ $1.84_{-0.29}^{+0.18}$  &\ $1.69_{-0.50}^{+0.37}KeV$\\
\end{tabular}
\end{ruledtabular}
\end{table*}

\subsection*{3.2 NUMERICAL RESULTS OF  THE LOCAL-QCDSR}

Now, the hadronic coupling constants $G_{\mathbb{TVP}}(Q^{2}=-M_{\mathbb{P}}^{2})$ and the corresponding decay widths of the local-QCDSR are obtained as well, using similar procedure mentioned above but only for the confirmation of the $Q^{2}$ intervals. The numerical results are listed in Table III.
\begin{table*}[tjhgg]
\begin{ruledtabular}\caption{The numerical results with the local-QCDSR }
\begin{tabular}{c c c c c c c c c c c c c c c c c c}
$p'^{\mu}\epsilon^{\nu\tau p p'}$($p'^{\nu}\epsilon^{\mu\tau p p'}$) & \  $ Q^{2}  (GeV^{2}) $ & \ $G_{\mathbb{TVP}}(GeV^{-2})$ &\ Decay Width $\Gamma_{i}$ \\
\hline
$ D_{2}^{*0} \rightarrow D^{*+}  \pi^{-} $  & \  $3.9\sim4.9$ & \ $3.65^{+1.18}_{-0.83}$  & \ $1.93^{+0.61}_{-0.78}MeV$  \\
$ D_{2}^{*0} \rightarrow D^{*0}  \pi^{0} $  & \  $3.9\sim4.9$ & \ $3.89^{+1.26}_{-0.89}$  & \ $1.16^{+0.87}_{-0.47}MeV$  \\
$ D_{s2}^{*+} \rightarrow D^{*0}  K^{+} $   & \ $3.9\sim4.9$ & \ $2.72^{+0.52}_{-0.49}$  & \ $99.71^{+43.10}_{-32.55}KeV$   \\
$ D_{s2}^{*+} \rightarrow D^{*+}  K^{0} $   & \  $3.9\sim4.9$ & \ $2.68^{+0.53}_{-0.47}$  & \ $74.83^{+32.27}_{-24.38}KeV$   \\
$ B_{2}^{*0} \rightarrow B^{*+}  \pi^{-} $  & \  $2.1\sim3.1$ & \ $2.91^{+0.26}_{-0.24}$  & \ $1.07^{+0.20}_{-0.16}MeV$  \\
$ B_{2}^{*0} \rightarrow B^{*0}  \pi^{0} $  & \  $2.1\sim3.1$ & \ $3.10^{+0.29}_{-0.25}$  & \ $0.62^{+0.12}_{-0.10}MeV$  \\
$ B_{s2}^{*0} \rightarrow B^{*+}  K^{-} $   & \ $2.1\sim3.1$ & \ $2.17^{+0.01}_{-0.09}$  & \ $4.41^{+0.01}_{-0.37}KeV$  \\
$ B_{s2}^{*0} \rightarrow B^{*0}  \bar{K}^{0} $   & \  $2.1\sim3.1$ & \ $2.14^{+0.01}_{-0.08}$  & \ $2.31^{+0.01}_{-0.25}KeV$  \\
\hline
$p^{\mu}\epsilon^{\nu\tau p p'}$($p^{\nu}\epsilon^{\mu\tau p p'}$) & \  $ Q^{2}  (GeV^{2}) $ & \ $G_{\mathbb{TVP}}(GeV^{-2})$ &\ Decay Width $\Gamma_{i}$  \\
\hline
$ D_{2}^{*0} \rightarrow D^{*+}  \pi^{-} $  & \  $3.9\sim4.9$ & \ $0.66^{+0.02}_{-0.04}$  & \ $63.96^{+6.79}_{-3.37}KeV$  \\
$ D_{2}^{*0} \rightarrow D^{*0}  \pi^{0} $  & \  $3.9\sim4.9$ & \ $0.72^{+0.02}_{-0.03}$  & \ $39.58^{+1.59}_{-2.81}KeV$  \\
$ D_{s2}^{*+} \rightarrow D^{*0}  K^{+} $   & \ $3.9\sim4.9$ & \ $0.72^{+0.00}_{-0.05}$  & \ $6.99^{+0.00}_{-0.89}KeV$   \\
$ D_{s2}^{*+} \rightarrow D^{*+}  K^{0} $   & \  $3.9\sim4.9$ & \ $0.70^{+0.00}_{-0.04}$  & \ $5.16^{+0.02}_{-0.61}KeV$   \\
\end{tabular}
\end{ruledtabular}
\end{table*}

\subsection*{3.3 DISCUSSION ON THE RESULTS}

As shown in Table II and III, there are the numerical results associate with the structures $p'^{\mu}\epsilon^{\nu\tau pp'}$ and $p'^{\nu}\epsilon^{\mu\tau pp'}$ for both the QCDSR and the local-QCDSR. Meanwhile, the structures $p^{\mu}\epsilon^{\nu\tau pp'}$ and $p^{\nu}\epsilon^{\mu\tau pp'}$ are useful only for decays $D_{2}^{*}(2460)^{0}\rightarrow D^{*}\pi$ and $D_{s2}^{*}(2573)\rightarrow D^{*}K$ with the local-QCDSR.
 These 4 structures can be considered as 4 decay modes (or 'channels'). So, the decay widths $\Gamma$ of every processes should be the sum of the decay width associate with the individual structure, i.e. $\Gamma=\sum \Gamma_{i}$. For the convenience of discussion, the decay widths $\Gamma$ of every processes are defined as the partial widths $\Gamma$ of the tensor mesons. Now, the numerical results of the partial widths $\Gamma$ of every tensor mesons are listed in Table IV. It is shown that the numerical results in this work with two approaches (the QCDSR and the local-QCDSR) are closed to each other (see the former three columns in Table IV). This indicates that the local approximation approach is reasonable, and the vacuum condensates contribution are very small for the strong decay of the heavy tensor mesons because the $\langle GG\rangle$ and $\langle GGG\rangle$ condensates actually vanish in calculations with the local-QCDSR.

\begin{table*}[tjhgg]
\begin{ruledtabular}\caption{The partial widths $\Gamma$ of every tensor mesons. }
\begin{tabular}{c c c c c c c c c c c c c c c c c c}
Processes (this work) & \  $\Gamma$ (the QCDSR) & \ $\Gamma$ (the local-QCDSR) &\ Processes~\cite{Wang2} &\ $\Gamma$ (the local-QCDSR)~\cite{Wang2}  \\
\hline
$D_{2}^{*0}\rightarrow D^{*+}\pi^{-}$ &\ $3.94^{+2.46}_{-1.50}MeV$ &\ $3.99^{+1.22}_{-1.56}MeV$ &\ $ D_{2}^{*0} \rightarrow D^{+}\pi^{-} $ &\ $7.91^{+3.49}_{-3.00}MeV$  \\
$ D_{2}^{*0} \rightarrow D^{*0}  \pi^{0} $  & \  $2.36^{+1.48}_{-0.90}MeV$ & \ $2.40^{+1.74}_{-0.94}MeV$  & \ $ D_{2}^{*0} \rightarrow D^{0}  \pi^{0} $  & \ $4.14^{+1.82}_{-1.57}MeV$  \\
$ D_{s2}^{*+} \rightarrow D^{*0}  K^{+} $   & \ $0.21^{+0.07}_{-0.08}MeV$ & \ $0.20^{+0.09}_{-0.07}MeV$  & \ $ D_{s2}^{*+} \rightarrow D^{0}  K^{+} $   & \ $3.35^{+1.48}_{-1.27}MeV$   \\
$ D_{s2}^{*+} \rightarrow D^{*+}  K^{0} $   & \  $0.16^{+0.05}_{-0.05}MeV$ & \ $0.15^{+0.06}_{-0.05}MeV$  & \ $ D_{s2}^{*+} \rightarrow D^{+}  K^{0} $   & \ $3.04^{+1.34}_{-1.15}MeV$   \\
$ B_{2}^{*0} \rightarrow B^{*+}  \pi^{-} $  & \  $1.84^{+0.04}_{-0.12}MeV$ & \ $2.14^{+0.40}_{-0.32}MeV$  & \ $ B_{2}^{*0} \rightarrow B^{+}  \pi^{-} $  & \ $3.42^{+0.90}_{-0.85}MeV$  \\
$ B_{2}^{*0} \rightarrow B^{*0}  \pi^{0} $  & \  $1.04^{+0.04}_{-0.06}MeV$ & \ $1.24^{+0.24}_{-0.20}MeV$  & \ $ B_{2}^{*0} \rightarrow B^{0}  \pi^{0} $  & \ $1.73^{+0.46}_{-0.43}MeV$  \\
$ B_{s2}^{*0} \rightarrow B^{*+}  K^{-} $   & \ $6.46^{+1.42}_{-1.88}KeV$ & \ $8.82^{+0.02}_{-0.74}KeV$  & \ $ B_{s2}^{*0} \rightarrow B^{+}  K^{-} $   & \ $0.25^{+0.06}_{-0.06}MeV$  \\
$ B_{s2}^{*0} \rightarrow B^{*0}  \bar{K}^{0} $   & \  $3.38^{+0.74}_{-1.00}KeV$ & \ $4.62^{+0.02}_{-0.50}KeV$  & \ $ B_{s2}^{*0} \rightarrow B^{0} \bar{K}^{0} $   & \ $0.21^{+0.06}_{-0.05}MeV$  \\

\end{tabular}
\end{ruledtabular}
\end{table*}

In addition, the following relations obtained from the experimental data~\cite{Ber} are used in this work.
\begin{eqnarray}
&\frac{\Gamma_{exp.}(D_{2}^{*}(2460)^{0}\rightarrow D^{+}\pi^{-})}{\Gamma_{exp.}(D_{2}^{*}(2460)^{0}\rightarrow D^{*+}\pi^{-})}= 1.54\pm 0.15, \notag \\
&\frac{\Gamma_{exp.}(B_{2}^{*}(5747)^{0}\rightarrow B^{+}\pi^{-})}{\Gamma_{exp.}(B_{2}^{*}(5747)^{0}\rightarrow B^{*+}\pi^{-})}= 0.91\pm 0.38\pm 0.28 .
 \end{eqnarray}
According to the numerical results in Table IV, we give the following comparisons with the local-QCDSR.
\begin{eqnarray}
&\frac{\Gamma(D_{2}^{*}(2460)^{0}\rightarrow D^{+}\pi^{-})}{\Gamma(D_{2}^{*}(2460)^{0}\rightarrow D^{*+}\pi^{-})}= 1.98^{+1.06}_{-1.08}, ~~
   \frac{\Gamma(D_{2}^{*}(2460)^{0}\rightarrow D^{0}\pi^{0})}{\Gamma(D_{2}^{*}(2460)^{0}\rightarrow D^{*0}\pi^{0})}= 1.73^{+1.46}_{-0.94}, \notag \\
  & \frac{\Gamma(D_{s2}^{*}(2573)^{+}\rightarrow D^{0}K^{+})}{\Gamma(D_{s2}^{*}(2573)^{+}\rightarrow D^{*0}K^{+})}= 16.75^{+10.56}_{-8.64}, ~~
   \frac{\Gamma(D_{s2}^{*}(2573)^{+}\rightarrow D^{+}K^{0})}{\Gamma(D_{s2}^{*}(2573)^{+}\rightarrow D^{*+}K^{0})}= 20.27^{+12.06}_{-10.22}, \notag \\
   &\frac{\Gamma(B_{2}^{*}(5747)^{0}\rightarrow B^{+}\pi^{-})}{\Gamma(B_{2}^{*}(5747)^{0}\rightarrow B^{*+}\pi^{-})}= 1.59^{+0.51}_{-0.46}, ~~
   \frac{\Gamma(B_{2}^{*}(5747)^{0}\rightarrow B^{0}\pi^{0})}{\Gamma(B_{2}^{*}(5747)^{0}\rightarrow B^{*0}\pi^{0})}= 1.39^{+0.46}_{-0.41}, \notag \\
  & \frac{\Gamma(B_{s2}^{*}(5840)^{0}\rightarrow B^{+}K^{-})}{\Gamma(B_{s2}^{*}(5840)^{0}\rightarrow B^{*+}K^{-})}= 28.34^{+6.80}_{-7.21}, ~~
   \frac{\Gamma(B_{s2}^{*}(5840)^{0}\rightarrow B^{0}\bar{K}^{0})}{\Gamma(B_{s2}^{*}(5840)^{0}\rightarrow B^{*0}\bar{K}^{0})}= 45.45^{+0.20}_{-4.91}.
\end{eqnarray}
As shown in Eqs.(22) and (23), there are certain deviations between the calculated values and the experimental values for  $\frac{\Gamma(D_{2}^{*}(2460)^{0}\rightarrow D^{+}\pi^{-})}{\Gamma(D_{2}^{*}(2460)^{0}\rightarrow D^{*+}\pi^{-})}$ and $\frac{\Gamma(B_{2}^{*}(5747)^{0}\rightarrow B^{+}\pi^{-})}{\Gamma(B_{2}^{*}(5747)^{0}\rightarrow B^{*+}\pi^{-})}$. But the calculated values are close to the experimental values within the error ranges.
On the other hand, we find the decay processes in which the $s$ quark participates in our present work are suppressed largely about one order of magnitude compared with the partial widths in Ref.~\cite{Wang2}. As we know, it belongs to the kinematical suppression.

Finally, we can give the estimation of the full widths of $D_{2}^{*}(2460)^{0}$, $D_{s2}^{*}(2573)$,  $B_{2}^{*}(5747)^{0}$, and $B_{s2}^{*}(5840)^{0}$. In our calculations (this work and Ref.~\cite{Wang2}), all of the input parameters are taken from the experimental values, except for the decay constants $f_{D^{*}}$, $f_{B^{*}}$, $f_{D_{2}^{*}}$, $f_{D_{s2}^{*}}$, $f_{B_{2}^{*}}$ and $f_{B_{s2}^{*}}$. The values of $f_{D^{*}}$ and $f_{B^{*}}$ are taken from Ref.~\cite{wang3}, where the perturbative $\mathcal{O}(\alpha_{s})$ corrections have been considered. The tensor mesons' decay constants $f_{D_{2}^{*}}$, $f_{D_{s2}^{*}}$, $f_{B_{2}^{*}}$ and $f_{B_{s2}^{*}}$ are taken from Ref.~\cite{Wang1}, where the perturbative $\mathcal{O}(\alpha_{s})$ corrections are neglected originally and discussed later. In our calculations, we have taken the values of  $f_{D_{2}^{*}}$, $f_{D_{s2}^{*}}$, $f_{B_{2}^{*}}$ and $f_{B_{s2}^{*}}$ without corrections. So, we should estimate their values including the perturbative $\mathcal{O}(\alpha_{s})$ corrections. In the massless limits, taking into accounting the $\mathcal{O}(\alpha_{s})$ corrections amounts to multiplying the parturbative terms by a factor $(1-\frac{\alpha_{s}}{\pi})$~\cite{Reind82}, where the values of the strong running coupling constant $\alpha_{s}$ is the function of the energy scale $\mu$~\cite{Wang1,Wang2,wang3}. So, we estimate the values of the tensor mesons' decay constants by multiplying this correction factor.
Now, as to these 4 tensor mesons, the partial widths $\Gamma$ associate with the major decay processes are all listed in Table IV. Therefore, just adding up the partial widths of each menson and taking into account the perturbative $\mathcal{O}(\alpha_{s})$ corrections,  we obtain the full widths as follows:
 \begin{eqnarray}
   &G_{\mathbb{TDP}}\rightarrow G_{\mathbb{TDP}}/(1-\frac{\alpha_{s}}{\pi}),  \notag \\
    &G_{\mathbb{TVP}}\rightarrow G_{\mathbb{TVP}}/(1-\frac{\alpha_{s}}{\pi}),  \notag \\
   &\Gamma(D_{2}^{*}(2460)^{0})\rightarrow (14\sim 33) MeV ,  \notag \\
   &\Gamma(D_{s2}^{*}(2573))\rightarrow (5\sim 12) MeV ,  \notag \\
   &\Gamma(B_{2}^{*}(5747)^{0})\rightarrow (10\sim 16) MeV ,  \notag \\
   &\Gamma(B_{s2}^{*}(5840)^{0})\rightarrow (0.4\sim 0.6) MeV,
 \end{eqnarray}
where the hadronic coupling constant $G_{\mathbb{TDP}}$ comes from Ref.~\cite{Wang2}. Correspondingly, the relevant experimental data~\cite{KAO} are listed below,
\begin{eqnarray}
   &\Gamma_{exp.}(D_{2}^{*}(2460)^{0}) = (49.0\pm1.3)MeV,  \notag \\
   &\Gamma_{exp.}(D_{s2}^{*}(2573))= (17\pm4) MeV,  \notag \\
   &\Gamma_{exp.}(B_{2}^{*}(5747)^{0})=(23^{+\ 5}_{-11}) MeV,  \notag \\
   &\Gamma_{exp.}(B_{s2}^{*}(5840)^{0})=(1.6\pm0.5) MeV.
 \end{eqnarray}
 The theoretical value in (24) is still smaller than the experimental value. However, the predicted upper bound of $\Gamma(D_{s2}^{*}(2573))$ and $\Gamma(B_{2}^{*}(5747)^{0})$ is very closed to the lower bound of the  experimental data. Actually, the phase space of decays $D_{s2}^{*}(2573)\rightarrow D^{*}K$ and $B_{s2}^{*}(5840)^{0}\rightarrow B^{*}K$ analyzed in this work is suppressed, compared with that analyzed in Ref.~\cite{Wang2}. So, their contributions to the full widths are very small. However, the analysis of these processes is necessary to recognize the decay of the tensor mesons.

\section*{IV. CONCLUSIONS }

In this paper, in order to discuss the strong decay
 $D_{2}^{*}(2460)^{0}\rightarrow D^{*}\pi$, $D_{s2}^{*}(2573)\rightarrow D^{*}K$, $B_{2}^{*}(5747)^{0}\rightarrow B^{*}\pi$, and $B_{s2}^{*}(5840)^{0}\rightarrow B^{*}K$, we analyze the $TVP$ type of vertex, calculate the hadronic coupling constants $G_{D_{2}^{*}D^{*}\pi}$, $G_{D_{s2}^{*}D^{*}K}$, $G_{B_{2}^{*}B^{*}\pi}$ and $G_{B_{s2}^{*}B^{*}K}$ with the QCDSR and the local-QCDSR, and obtain the corresponding decay widths. The results show that the QCDSR
 and the local-QCDSR give the similar numerical results, which indicates these two approaches are both good at describing the heavy tensor mesons' strong decay in present work. These hadronic coupling conatants can be taken as basic input parameters in phenomenological analysis and the predicted partial widths could be meaningful to the relevant experiments in the futures. Finally, combined with the related results in Ref.~\cite{Wang2}, the full widths of $\Gamma(D_{2}^{*}(2460)^{0})$, $\Gamma(D_{s2}^{*}(2573))$, $\Gamma(B_{2}^{*}(5747)^{0})$ and $\Gamma(B_{s2}^{*}(5840)^{0})$ are obtained and compared with the experimental data.

\label{sec4}

\section*{ACKNOWLEDGMENTS}

This work is supported by National Natural Science Foundation of China, Grant Number 11375063, the Fundamental
Research Funds for the Central Universities, Grant Number 13QN59, 2014ZD42, and the Department of Science in Guizhou Province, Grant Number 2013GZ62432 and 2013GZ56464.

\label{sec4}

\section*{Appendix A}
In this appendix, we list the detailed results of the correlation function of $\langle GG\rangle$ and $\langle GGG\rangle$  condensates in the OPE side, after performing the double Borel transformation.
\begin{equation}
\begin{aligned}
&\hat {\Pi }_i^{ \langle GG \rangle } (M_1^2 ,M_2^2 ,Q^2)
= \hat {\Pi }_i^{gg - 1} +
\hat {\Pi }_i^{gg - 2} + \hat {\Pi }_i^{gg - 3} + \hat {\Pi }_i^{gg - 4} +\hat {\Pi }_i^{gg - 5} + \hat {\Pi }_i^{gg - 6} \\
&=\frac{1}{12\pi ^2}\left\langle {\frac{\alpha _s GG}{\pi }}
\right\rangle \mbox{\{} 6m_Q I_{\alpha-\beta}(00113) + 6(m_Q -
2m_q )m_Q ^2 I_{\alpha-\beta} (00114) \\
&+ 6(m_Q - 2m_q )m_{q'} ^2 I_{\alpha-\beta}(00141) - 2(m_Q - 2m_q )I_{\alpha-\beta}(00212) - 6(12m_q - 2m_Q ) I_{\alpha-\beta}(00221)\\
&  - 2(2m_q - 2m_Q )I_{\alpha-\beta}(00122) - 12m_q I_{\alpha-\beta}(00311) + 6(m_Q - 2m_q )m_q ^2 I_{\alpha-\beta}(00141) \} \notag\\
\end{aligned}
\end{equation}

\begin{equation}
\begin{aligned}
&\hat {\Pi }_j^{ \langle GG \rangle } (M_1^2 ,M_2^2 ,Q^2) = \hat {\Pi }_j^{gg - 1} +
\hat {\Pi }_j^{gg - 2} + \hat {\Pi }_j^{gg - 3} + \hat {\Pi }_j^{gg - 4} +
\hat {\Pi }_j^{gg - 5} + \hat {\Pi }_j^{gg - 6} \\
&=\frac{1}{12\pi ^2}\left\langle {\frac{\alpha _s GG}{\pi }}
\right\rangle \{ + 3m_Q I_\alpha (00113) + [3(m_Q - m_{q'} )m_{Q}^{2} +
6m_q m_Q ^2]I_\alpha (00114) \notag\\
& - 3m_{q'} I_\alpha (00131) + [6m_q m_{q'} ^2 - 3(m_{q'} - m_Q )m_{q'}
^2]I_\alpha (00141)+ \frac{1}{2}[6m_q + (m_Q - 3m_{q'} )]I_\alpha (00221) \notag\\
& + 6m_q I_\alpha (00311) + [6m_q ^3 - 3(m_{q'} - m_Q )m_q ^2]I_\alpha
(00411) + \frac{1}{2}(m_{q'} - m_Q - 2m_q )I_\alpha (00122) \\
& + \frac{1}{2}[ - 2m_q - (m_Q - m_{q'} )]I_\alpha (00212)\}
\end{aligned}
\end{equation}

\begin{equation}
\begin{aligned}
&\hat {\Pi }_i^{ \langle GGG \rangle } (M_1^2 ,M_2^2 ,Q^2)\\
&= \hat {\Pi }_i^{ggg - 1} +
\hat {\Pi }_i^{ggg - 2} + \hat {\Pi }_i^{ggg - 3} + \hat {\Pi }_i^{ggg - 4}
+ \hat {\Pi }_i^{ggg - 5} + \hat {\Pi }_i^{ggg - 6}
 + \hat {\Pi }_i^{ggg - 7} + \hat {\Pi }_i^{ggg - 8} + \hat {\Pi }_i^{ggg -
9} + \hat {\Pi }_i^{ggg - 10}  \notag\\
&= \frac{\left\langle {g_s ^3G^cG^aG^bf^{abc}}
\right\rangle }{12 \cdot (2\pi )^4}
 \{ - 4m_Q I_{\alpha-\beta}(10214)- (6m_Q + 8m_q )I_{\alpha-\beta}(00213)  + 4(2m_q - m_Q )m_Q ^2 I_{\alpha-\beta}(00214)  \notag\\
 &+ (6m_Q - 4m_q )I_{\alpha-\beta}(00123)  + (4m_Q - 2m_q )m_Q ^2I_{\alpha-\beta}
(00124)
 - 8m_Q I_{\alpha-\beta}(01124)- 4m_q I_\alpha (10124)  \notag\\
  &  + 4m_q I_{\alpha-\beta}(10412)  + 4(m_Q + 4m_q )m_q ^2 I_{\alpha-\beta}(00412)
 - 4m_q I_{\alpha-\beta}(01412) - 8m_Q m_q ^2 I_{\alpha-\beta}(00421) \notag\\
&  + 4(m_Q + 10m_q ) I_{\alpha-\beta}(00321)
  - (6m_Q - 24m_q + 8m_{q'} )I_{\alpha-\beta}(00231)
 - (2m_Q - 12m_q )m_{q'} ^2 I_{\alpha-\beta}(00241) \notag\\
  & - 8m_{q'}  I_{\alpha-\beta}(00141)
 - 4(m_Q - m_q ) I_{\alpha-\beta}(00132) - 4(m_Q - m_q )m_{q'} ^2 I_{\alpha-\beta}
(00142) - 8m_{q'}  I_{\alpha-\beta}(01142)\notag\\
 & - 4m_{q'} I_\alpha (10142)
 + 12(3m_Q - m_q )I_{\alpha-\beta}(00114)  + 3(10m_Q -
32m_q )m_Q ^2 I_{\alpha-\beta}(00115) +30m_{q'} I_\alpha (10151)\notag\\
&+ 2(m_Q - 2m_q ) I_{\alpha-\beta}(00222)
+3(2m_Q - 24m_q ) I_{\alpha-\beta}(00411) -16(m_Q - m_q )m_q
^2 I_{\alpha-\beta}(00511) \}
\end{aligned}
\end{equation}

\begin{equation}
\begin{aligned}
&\hat {\Pi }_j^{ \langle GGG \rangle } (M_1^2 ,M_2^2 ,Q^2)\\
&= \hat {\Pi }_j^{ggg - 1} +
\hat {\Pi }_j^{ggg - 2} + \hat {\Pi }_j^{ggg - 3} + \hat {\Pi }_j^{ggg - 4}
+ \hat {\Pi }_j^{ggg - 5} + \hat {\Pi }_j^{ggg - 6}
 + \hat {\Pi }_j^{ggg - 7} + \hat {\Pi }_j^{ggg - 8} + \hat {\Pi }_j^{ggg -
9} + \hat {\Pi }_j^{ggg - 10} \notag\\
&= \frac{1}{12 \cdot (2\pi )^4}\left\langle {g_s ^3G^cG^aG^bf^{abc}}
\right\rangle
 \{(4m_q  - 2m_{q'} +2 m_Q )m_Q ^2 I_\alpha (00124) + (2m_q -
3m_Q )I_\alpha (00123) \notag\\
& - (m_{q'} + 3m_Q + 4m_q )I_\alpha (00213) -  2(2m_q + m_{q'} + m_Q )m_Q ^2 I_\alpha (00214)+
2m_Q I_\alpha (01214)
 + 4m_Q I_\alpha (00114)\notag\\
 & - 4m_Q I_\alpha (10214)
 - 2m_q I_\alpha (00411) - 2(4m_q -m_Q + m_{q'} )m_q ^2 I_\alpha
(00412) - (m_{q'} - 2m_q )I_\alpha (00312) \notag\\
& + (2m_Q - 24m_q )I_\alpha (00321) - 4m_Q m_q ^2 I_\alpha (00421) - 4m_q
I_\alpha (00411) + (9m_{q'} - 12m_q - 3m_Q )I_\alpha (00231)\notag\\
& - (m_Q - 12m_q )m_{q'}
^2I_\alpha (00241) + 4m_{q'} I_\alpha (00141)+ 8m_{q'} I_\alpha (01142)
 - 2(m_{q'} - m_Q )m_{q'} ^2I_\alpha (00142) \notag\\
& +  4m_{q'} I_\alpha (10142) + (2m_q - 3m_{q'} + 2m_Q )I_\alpha
(00132) + 3(6m_Q - m_{q'} + 2m_q )I_\alpha (00114)+ 60m_{q'} I_\alpha (01151) \notag\\
 &+ 3(5m_Q - 8m_{q'} - 16m_q
)m_Q ^2I_\alpha (00115) + 3( 12m_q + m_Q - m_{q'} )I_\alpha (00411) +24(m_Q - m_{q'} )m_q
^2I_\alpha (00511) \notag\\
& + (2m_q - m_{q'} + m_Q )I_{\alpha} (00222)\}
\end{aligned}
\end{equation}

\section*{Appendix B}
In this appendix, some formulas used in Appendix A are listed below.
\begin{equation}
 I_{\alpha-\beta}(MNabc)\equiv  I_{\alpha}(MNabc)- I_{\beta}(MNabc)\notag\\
\end{equation}
\begin{equation}
\begin{aligned}
&I_{\kappa}(00abc)=\frac{(-1)^{a+b+c+\delta}\pi^{2}}{\Gamma(a)\Gamma(b)\Gamma(c)(M^{2})^{c-2-\delta}(M_{2}^{2})^{a-1}(M_{1}^{2})^{b-1+\delta}}
\int_{0}^{1}d\lambda'\frac{\lambda'^{4-a-b+\delta}}{(1-\lambda')^{c-1-\delta}}
F(\lambda', M_{1}^{2},M_{2}^{2},Q^{2})\notag\\
\end{aligned}
\end{equation}
\begin{equation}
\begin{aligned}
I_{\kappa}(01abc)&=\frac{(-1)^{a+b+c+\delta}\pi^{2}[-(a-2)]}{\Gamma(a)\Gamma(b)\Gamma(c)(M^{2})^{c-2-\delta}(M_{2}^{2})^{a-2}(M_{1}^{2})^{b-1+\delta}}
\int_{0}^{1}d\lambda'\frac{\lambda'^{4-a-b-\delta}}{(1-\lambda')^{c-1-\delta}}
F(\lambda', M_{1}^{2},M_{2}^{2},Q^{2})\notag\\
&+\frac{(-1)^{a+b+c+\delta}\pi^{2}[-(c-3)]}{\Gamma(a)\Gamma(b)\Gamma(c)(M^{2})^{c-3-\delta}(M_{2}^{2})^{a-1}(M_{1}^{2})^{b-1+\delta}}
\int_{0}^{1}d\lambda'\frac{\lambda'^{4-a-b-\delta}}{(1-\lambda')^{c-1-\delta}}
F(\lambda', M_{1}^{2},M_{2}^{2},Q^{2})\notag\\
&+\frac{(-1)^{a+b+c+\delta}\pi^{2}}{\Gamma(a)\Gamma(b)\Gamma(c)(M^{2})^{c-2-\delta}(M_{2}^{2})^{a-2}(M_{1}^{2})^{b-1+\delta}} \notag\\
&\int_{0}^{1}d\lambda'\frac{\lambda'^{4-a-b-\delta}}{(1-\lambda')^{c-1-\delta}}
\{-\frac{(1-\lambda')Q^{2}M_{2}^{2}}{\lambda'(M_{1}^{2}+M_{2}^{2})^{2}}+\frac{m_{q}^{2}}{\lambda' M_{2}^{2}}+\frac{m_{Q}^{2}}{(1-\lambda')M_{2}^{2}}\}
F(\lambda', M_{1}^{2},M_{2}^{2},Q^{2})\notag\\
\end{aligned}
\end{equation}
\begin{equation}
\begin{aligned}
I_{\kappa}(10abc)&=\frac{(-1)^{a+b+c+\delta}\pi^{2}[-(b-1)]}{\Gamma(a)\Gamma(b)\Gamma(c)(M^{2})^{c-2-\delta}(M_{2}^{2})^{a-1}(M_{1}^{2})^{b-2+\delta}}
\int_{0}^{1}d\lambda'\frac{\lambda'^{4-a-b-\delta}}{(1-\lambda')^{c-1-\delta}}
F(\lambda', M_{1}^{2},M_{2}^{2},Q^{2})\notag\\
&+\frac{(-1)^{a+b+c+\delta}\pi^{2}[-(c-3)]}{\Gamma(a)\Gamma(b)\Gamma(c)(M^{2})^{c-3-\delta}(M_{2}^{2})^{a-1}(M_{1}^{2})^{b-1+\delta}}
\int_{0}^{1}d\lambda'\frac{\lambda'^{4-a-b-\delta}}{(1-\lambda')^{c-1-\delta}}
F(\lambda', M_{1}^{2},M_{2}^{2},Q^{2})\notag\\
&+\frac{(-1)^{a+b+c+\delta}\pi^{2}}{\Gamma(a)\Gamma(b)\Gamma(c)(M^{2})^{c-2-\delta}(M_{2}^{2})^{a-1}(M_{1}^{2})^{b-2+\delta}} \notag\\
&\int_{0}^{1}d\lambda'\frac{\lambda'^{4-a-b-\delta}}{(1-\lambda')^{c-1-\delta}}
\{+\frac{(1-\lambda')Q^{2}M_{1}^{2}}{\lambda'(M_{1}^{2}+M_{2}^{2})^{2}}+\frac{m_{q'}^{2}}{\lambda' M_{1}^{2}}+\frac{m_{Q}^{2}}{(1-\lambda')M_{1}^{2}}\}
F(\lambda', M_{1}^{2},M_{2}^{2},Q^{2})\notag\\
\end{aligned}
\end{equation}
with $\delta=1$, if $\kappa=\alpha$, or $\delta=0$, if $\kappa=\beta$, $F(\lambda', M_{1}^{2},M_{2}^{2},Q^{2})$ and $M^{2}$ are defined as
\begin{equation}
 F(\lambda', M_{1}^{2},M_{2}^{2},Q^{2})=Exp\{-\frac{(1-\lambda')Q^{2}}{\lambda'(M_{1}^{2}+M_{2}^{2})}-\frac{m_{q}^{2}}{\lambda' M_{2}^{2}}
-\frac{m_{q'}^{2}}{\lambda' M_{1}^{2}}-\frac{m_{Q}^{2}}{(1-\lambda') M^{2}}\}\notag\\
\end{equation}
\begin{equation}
 \frac{1}{M^{2}}=\frac{1}{M_{1}^{2}}+\frac{1}{M_{2}^{2}}\notag\\
\end{equation}


\end{document}